Title: Energetic frustrations in protein folding at residue resolution: a simulation study of homologous immunoglobulin-like β-sandwich proteins

Running title: Energetic frustrations in β-sandwich proteins

Authors: Yunxiang Sun[1] and Dengming Ming[*]

College of Biotechnology and Pharmaceutical Engineering, Nanjing Tech University, 30 Puzhu South Road, Nanjing, Jiangsu 211816, PR China

[*]Contact information:
Dengming Ming
Biotech Building Room B1-404
College of Biotechnology and Pharmaceutical Engineering,
Nanjing Tech University,
30 South Puzhu Road
Jiangsu 211816, PR China
Tel: 8625-58139942
Email: dming@njtech.edu.cn

[1]Now in Department of Physics and Astronomy, Clemson University, Clemson, South Carolina 29634, USA

**Abstract**

Nonnative residual interactions have attracted increasing attention in recent protein folding researches. Experimental and theoretical investigations had been set out to catch nonnative contacts that might dominate key events in protein folding. However, energetic frustrations caused by nonnative inter-residue interactions are not systematically characterized, due to the complicated folding mechanism. Recently, we studied the folding of a set of homologous all-$\alpha$ proteins and found that nonnative-contact-based energetic frustrations are closely related to protein native-contact networks. In this paper, we studied the folding of nine homologous immunoglobulin-like (Ig-like) β-sandwich proteins and examined energetic frustrations caused by nonnative inter-residue interactions, based on analyses of residual $\phi$-values and contact maps of transition state ensembles. The proteins share similar tertiary structures, thus minimize topology frustration differences in protein folding and highlighting the effects of energetic frustrations as caused by hydrophilic-hydrophobic mutations. Our calculations showed that energetic frustrations have highly heterogeneous effects on the folding of most secondary structures and on the folding correlations between different folding-patches of β-sandwich proteins. The simulations revealed a strong interplay between energetic frustrations and native-contact networks in β-sandwich domains, which ensures that β-sandwich domains follow a common folding mechanism. Our results suggested that the folding processes of β-sandwich proteins might be redesigned by carefully manipulating energetic frustrations at residue level.

# INTRODUCTION

Most nascent polypeptide chains undergo a series of folding events before they acquire the compact biologically functioning three-dimensional (3D) conformations. The dynamics of inter-residue contacts defines details of protein folding kinetics, and thus determines the final native structures. Accurate modeling residue interactions in generic protein folding processes is still an ongoing challenge in the community, though much progress had been made both experimentally and theoretically[1]. Specifically, in protein folding some residues happen to collide, stay together and form the so-called native contacts in the native structures, while others only form transient contacts --- called non-native contacts that are separated in the final structures. Both experimental data and theoretical simulations have suggested that native-contacts play essential roles in determining both the protein folding kinetics and the native structures[2-5]. At the same time, minimally frustrated models of protein folding were developed, where non-native inter-residue interactions are systematically suppressed[6-9]. These models lead to the success of the free energy landscape theory in solving the kinetic paradox of protein folding problems [10-14], and in interpreting a variety of experimental studies such as single molecule pulling[15-17]. According to free-energy landscape theory, the formation of a few key native contacts in the polypeptide chain can start a series of down-hill like conformation changes towards the protein native states, without being trapped in any intermediate states at a local minimum.

Although native-contacts dominate the folding kinetics of many proteins, recent experiments and theoretical calculations found that non-native contacts also play a non-trivial role in protein folding[18-25]. Many non-native contacts are found in intermediate states of protein folding between the unfolded state and the fully folded native state. For example, non-native hydrogen bonding was identified between strand $\beta_1$ and strand $\beta_2$, stabilizing an intermediate state of a SH3 domain under native condition, using the Carr-Purcell-Meiboom-Gill (CPMG) relaxation dispersion nuclear magnetic resonance (RDNMR) spectroscopy technique[26]. The formation of nonnative contacts may either speed up the folding by stabilizing transition states and lowering the folding free energy barrier[18,27,28], or slow it down by trapping the molecules within local minimums in the free energy landscape[29]. Clearly, the detailed behavior of nonnative contacts is very sensitive to the sequence content of the protein, and its environment where folding happens. These observations inspired much interest in the community to develop computational models for the simulation of nonnative interactions in protein folding[30-39]. Non-native contacts can be built among different types of residues, of which those formed between hydrophobic residues, called nonnative hydrophobic interactions, were studied extensively. For example, recently Chan *et al.* found that the experimentally-observed intermediate state in the folding of the bacterial immunity protein Im7 can be modeled with the nonnative contacts at the hydrophobic core by the introduction of nonnative hydrophobic interactions to a minimally-frustrated model[36]. Another type of nonnative contact formed between charged residues, called nonnative salt bridge, is particularly important when protein folding takes place in certain liquid electrolyte such as aqueous urea solution[24].

Two types of frustrations for protein folding had been introduced as an extension to the minimal frustration principle --- topological frustration and energetic frustration[40]. Topological frustration results from the fact that the protein is a string of concatenated amino acid residues and holds particular 3D shape of native fold. The manner in which the amino acids are connected, and the residue's excluded volume prevent the protein from sampling certain area of configuration space, thus causing some parts of the protein less likely to contact one another, and hence resulting in folding frustration. For example, local contacts formed between immediately neighboring amino acids usually have a bigger chance to occur than do the distant contacts. Also, in terms of the overall protein shape, irregular topologies often have much more difficulties to fold than do the little symmetric ones[41], thus have more frustrating in folding. Energetic frustration is directly associated to the incorrect inter-residue interactions, where mismatched residues form contacts that are not presented in the native configurations. It is believed that natural selection helps to minimize energetic frustration by carefully choosing the amino acid sequence for the protein peptides. The mechanism of frustration has now become an important line of research in protein folding studies, which is usually investigated by analyzing the conformation distribution in the so-called protein transition state ensembles or TSEs. In terms of free energy landscape theory, TSEs are characterized by the profile maxima on reaction paths that connect the fully folded native-like states and the unfolded states in protein folding[42]. For example, Sutto *et al.* identified local frustrations, mostly energetic frustrations, in the on-pathway intermediates of the four-helix Im7 protein using an all-atom AMW model[31]. Shea *et al.* analyzed protein folding transition states to study the degree of energetic frustration in the folding of a 46-mer four standard β-barrel model protein[43] and the topological frustration in the folding of a fragment B of protein A[30]. Zarrine-Afsar *et al*. used TSEs to study the effects of energetic frustrations imposed at specific sites on the folding energetics of the Fyn SH3 domain[35]. Hills *et al*. studied transition states in α/β/α sandwich CheY-like proteins, and showed evidence that topological frustrations may arise from the competition of local folding in symmetric parts of the protein[44]. More interestingly, Contessoto *et al*. studied the interplay between energetic and topological frustrations for a set of 19 proteins, and identified distinct effects of frustrations on the folding rates depending on the topologies to studied proteins[45]. Very recently, Chung *et al*. combined single-molecule fluorescence with all-atom molecular dynamics simulations to study energetic frustrations in a designed α-helical protein, and identified specific inter-residue contacts that decreased the folding rate[24]. In summary, the effects of frustrations on protein folding are very complicated and involve many structural and dynamic factors due to changes in topologies and amino acid sequences of studied proteins.

Recently we studied the energetic frustration mechanism in folding simulations of five homologous four-helical Im9 proteins[46]. The selected proteins share similar 3D topological structures, but have a single hydrophilic-to-hydrophobic mutual mutation. Hence, these proteins have the same native contact network with only single mutation scattered at a single node of the network. We noticed that although similar comparative folding simulations had been performed to illustrate the common folding mechanism for homologous proteins[47,48], no residual level frustrations had been analyzed through comparative studies of homologous proteins. Our studies probed energetic frustrations at residual level for the folding of all-α proteins, and revealed that energetic frustrations are very sensitive to the local structural environments where they are introduced. Specifically, we found that frustrations at regions of dense native contacts cause large perturbation on the folding of all-α proteins. The importance of local native contact density was also emphasized in a recent protein folding study of bacterial Immunity proteins Im7/Im9 whose secondary structure component are also all α-helices[36] --- there the interplay between local native contact density and frustration-relevant hydrophobicity is likely responsible for the big difference in folding kinetics between Im7 and Im9. Considering the fundamental difference of the 3D topologies between all-β sandwich proteins and all-α proteins, it is interesting to (1) explore the effects of energetic frustrations on the folding of all-*β* proteins, and (2) investigate the interplay between the energetic frustration and the local native contact density in all-*β* proteins.

In this paper, we examined energetic frustrations at residue level via *ϕ*-value analysis of TSEs, of five homologous β-sandwich immunoglobulin-like proteins and additional four artificially mutated structures, using a similar strategy as that in reference[46]. The all-*β* sandwich proteins were selected from the same domain entry in the structural classification of proteins (SCOP)[49], and shared similar 3D topologies and amino acid sequences. The structural similarity among these selected structures minimized the difference in their topological frustrations, thus enabling us to isolate the effects of energetic frustrations at specific locations where hydrophilic-via-hydrophobic mutations occur. Previous studies showed that, unlike all-α proteins, all-β proteins usually share similar folding pathways[48]. Hence, we might expect the behavior of energetic frustration in all-β protein to be quite different from that in the all-α proteins. Our calculations revealed that energetic frustrations are highly heterogeneous in the examined β-proteins, depending on the local environment where nonnative inter-residue contacts are introduced. Specifically, we found a larger distribution of energetic frustrations in the folding of the all-beta proteins, compared to the all-α proteins. We ascribed this difference to the particular topology of all-β proteins, where no native contact center exists and the local native contact density is more evenly distributed, compared to the all-α proteins. Our simulations have thus provided new insights to the effect of energetic frustrations in the folding of small all-β proteins.

**Methods**

*Homologous domains of Immunoglobulin-like beta-sandwich structures* The key of this study hinges on comparing the folding of homologous all-*β* proteins that share very similar 3D topology structures and amino acid sequences. Although Ig-like *β*-sandwich fold had been subjected to wide studies as a model of all-*β* system in protein folding[47,50-52], the examined proteins in published works vary a lot in either structure or sequences. We selected protein candidates from the immunoglobulin-like domain entry of b.1.1.1 in the SCOP database[49,53]. Five Ig-like *β*-sandwich domains d1gigl1, d3ks0l1, d4a6yl1, d2y06l1, d1etzl1 were selected, and they have almost identical amino acid sequence except a single dominant hydrophilic-to-hydrophobic mutual mutation in a total 110 residues (Figure 1). The mutations were recorded based on alignment with the protein domain d1gig1l (see caption of Figure 1). These domains share the same complex Greek-key topology, and their 3D structures are described as two tightly packed *β*-sheets packing (Figure 2a). The mutual RMSDs between selected structures are between 0.4 and 0.8Å, which minimized topological frustration due to the structure differences among studied proteins. Unlike the topology of all-α homologous Im9 domains where relatively dense native contacts exist in the center[54], here, in the β-sandwich structures, native contact distribution is fairly even. It is at the β-strand-connection areas where native contacts fluctuate relatively larger, and, in these areas, selected structure exhibit most of the mutual mutations (Figure 2b). Based on this observation, four additional single-site mutants were introduced artificially at different connection-areas so as to explore energetic frustration effects at these new positions. Table 1 summarized the difference between the 9 studied all-β structures and the mutations. For clarity, each examined domain is renamed after their representative mutation as listed in the caption of Figure 1.

*Energetic frustration model* The details of protein folding simulations had been described in the previous protein folding studies of Im9 all-α domains[54], here we only summarize a few key aspects of the calculations. The minimal frustrated protein folding was performed with an alpha carbon based coarse-grained Gō-like model, whose driving force is determined by the native topology of the studied proteins[55-60]. In Gō-like model the potential energy has the following form,

$$U = \sum_{i=1}^{N-1} \frac{K_b}{2}(r_i - r_{i,0})^2 + \sum_{i=1}^{N-2} \frac{K_\theta}{2}(\theta_i - \theta_{i,0})^2 + \sum_{i=1}^{N-3} \frac{K_\phi}{2}\left(1 - \cos\left(2\phi_i - \frac{\pi}{2}\right)\right)^2$$
$$+ \sum_{i<j-3}^{NC} \epsilon_{ij}\left[5\left(\frac{r_{ij,0}}{r_{ij}}\right)^{12} - 6\left(\frac{r_{ij,0}}{r_{ij}}\right)^{10}\right] + \sum_{i<j-3}^{NNC} \epsilon_{ij}\left(\frac{C}{r_{ij}}\right)^{12}$$

(1)

where $r, \theta, \phi$ represent instantaneous bond lengths, bond angles and dihedral angles respectively, and the subscript "0" stands for quantities measured in the native state configurations. The upper limit "*NC*" stands for native contacts and indicates summation over all inter-residue contacts found in native structures, and "NNC" for nonnative contacts and summation over inter-residue pairs that are not directly in contact in the native state. Two residues are determined to form native contact if their minimum atom distance is less than a cutoff value of 5.5Å. In a transition state, we define that a native contact holds their contact state only when the two $C_\alpha$ distance satisfies $r_{ij}/r_{ij,0} \leq 1.2$. Parameters for the model are $K_b = 200\epsilon_0 \text{Å}^{-2}$, $K_\theta = 40\epsilon_0 \text{rad}^{-2}$, $K_\phi = 0.3\epsilon_0$ and $C = 4\text{Å}$. An absolute energy value of $\epsilon_0 = 1.89 Kcal \cdot mol^{-1}$ was used following Ref[55] by assuming a folding temperature *T*=350K for protein G B1 domain. To distinguish different residue pairs, an MJ-flavored coefficient is assigned to a native contact so that the contact energy has the form of $\epsilon_{ij} = \epsilon_{ij}^{MJ} \epsilon_0$, where $\epsilon_{ij}^{MJ}$ is proportional to the knowledge-based inter-residue contact energies [61,62] and has been normalized by setting the averaged $\bar{\epsilon}_{ij} = 0.18\epsilon_0$. Now, energetic frustrations are introduced by nonnative hydrophobic inter-residue contacts to the minimal frustration model as following:

$$U_{\text{frustration}} = \sum_{i<j-3}^{NNC} \sigma_{ij}\epsilon_{ij} \left[ 5\left(\frac{C_f}{r_{ij}}\right)^{12} - 6\left(\frac{C_f}{r_{ij}}\right)^{10} \right] \quad (2)$$

where

$$\sigma_{ij} = \begin{cases} 0.5 & \text{if both residue } i \text{ and } j \text{ are hydrophobic,} \\ 0 & \text{else} \end{cases}$$

and $C_f = 5.5\text{Å}$. The protein folding simulations were then performed with the Langevin dynamics scheme, using a time step $\Delta t = 0.007\tau$ and a high friction coefficient $\beta = 0.2/\tau$. Here $\tau$ is the characteristic time of the system $\tau_0 = l_0(m/\varepsilon_0)^{1/2}$ which is 1.47ps when setting an averaged residue mass *m* of 119 a.m.u., and an averaged distance of $l_0 = 3.8$Å between adjacent $C_\alpha$ atoms. Choosing $\sigma = 0.5$ is a compromise between the folding efficiency (folding/unfolding transition events dramatically decrease with $\sigma$) and the folding frustration (which increases with $\sigma$). We noticed that recently a different Guassian type function was also used in the study of energetic frustrations in the Fyn SH3 domain folding[35].

*The variable temperature protein folding simulation* The protein folding simulations is usually carried out at the transition temperature $T_\theta$ at which the system spontaneously changes between the folded and unfold states. In practice, $T_\theta$ is determined as the collapse temperature when a maximum specific heat is reached as a function of simulation temperature[57,58]. However, finding specific heat maxima is not trivial, because it requires a large amount of long time equilibrium simulations at a series of trial temperatures. Further, the value of specific heat as measured by internal energy fluctuation is parameter-sensitive, and thus not easily determined accurately. To circumvent this difficulty, in a previous study, we developed a variable temperature folding (VTF) simulation method. The method is based on the observation that, at transition temperature, a fast-folding small protein may frequently shift between folded and unfolded states, and thus has equal opportunity to sample either of the two states. In the protein folding free energy landscape, this scenario is featured with a highly-raised barrier in the middle of two valleys, corresponding to the folded and unfolded states, respectively[30,43]. The VTF method skips the step of determining $T_\theta$, instead it continuously increases or decreases the simulation temperature, thus enforce the system to frequently change its states between folded and unfolded states, ensuring that the system has equal chance to sample either states. Specifically, in VIF calculations, Langevin dynamics simulation is firstly performed for $N_T$ steps with an initial guess collapse temperature $T$. Then the up-to-now trajectory is collected and its snapshots are grouped based on the number of the total native-contacts (or Q number) of conformation, and a probability density function (PDF) of native-contact number is determined using a histogram method. Usually, when $T$ is close to the true collapse temperature $T_\theta$, two peaks will appear in the PDF profile, of which one peak, called the non-native peak, corresponds to smaller Q and denotes the unfolded states, and the other one, called native peak, corresponds to larger Q. Next, the simulation temperature $T$ is slightly changed by a small value of $\Delta T$ as following: $T = T - \Delta T$ if the nonnative peak is higher than the native peak, or $T = T + \Delta T$ when the reverse applies ($T$ is randomly updated when the two peaks have the same height). This procedure is kept on going when the trajectory reaches a satisfactory length.

The validity of VTF had been carefully examined with the folding simulations of protein G B1 domain (PDB code 2GB1, 56 amino acids) and an all-α Im7 protein (SCOP domain d1ayia_, 86 amino acids). It was shown that VTF simulation give nearly identical free-energy profiles as does the conventional fixed-temperature simulation. The calculation results also suggested that ϕ-values (see the flowing) derived from VTF simulations are highly consistent with those derived from conventional simulations, marked by a Pearson correlation of 0.99 and a small standard derivation of 0.01~0.04. Compared with conventional fixed-temperature folding simulations, VTF needs only a single lengthy trajectory for folding simulation of one protein. In present studies, a typical simulation produces a trajectory that spans $3\mu s$ or has $3\times10^9$ steps. Here we set $\Delta T=0.002$ (which is about $0.5\%T_\theta$) and $N_T = 2\times10^7$, thus a typical trajectory includes 150 times of temperature updating.

*ϕ-value analysis and contact maps* In order to investigate the consequence of energetic frustrations in protein folding, we need to analyze the dynamic feature of the protein configuration in different thermodynamic states including the folded states, unfolded states and the transition states. The first quantity that characterizes protein states might be the total number of native contacts, the Q value, preserved in a given state. A statistical distribution of the state density with respect to Q describes the free energy landscape of the protein folding reactions, from which the folded and unfolded states are usually separated as two peaks in the landscape. Transition state ensembles or TSEs are usually defined as the groups of conformations sampled around free energy barrier that separate the folded and unfolded states. It is also equivalent to a collection of structures that locate in the valley between the two peaks[42,63-65]. The single Q-value, however, by itself poorly characterizes the TSEs, since for a given Q-value, the local native-contacts may have distinct distributions from one protein to another, and it is the detailed native-contact distribution that determines the protein folding mechanism. For this sake, a ϕ-value analysis was introduced to measures perturbations of local native-contacts at residue level in both experimental[64] and theoretical studies[66]. Essentially, ϕ-value compares the preserved native-contacts of a single residue in TSEs with those in unfolded and folded states. Here *ϕ-value* has the following form after Refs. [67,68],

$$\phi_i = \frac{\langle N_i \rangle^{\text{TSE}}}{N_i^{\text{nat}}} \tag{3}$$

where $N_i$ is the number of preserved native-contacts of the *i*th residue, the denominator is the number of native-contacts in the native state and the numerator the averaged number in TSEs. $\phi_i = 1$ when residue *i* has exactly the same number of native contacts in TSEs as does in the native state, indicating that TSEs preserve full folded state at this residue. $\phi_i = 0$ indicates TSEs fall in a fully unfolded state. *ϕ-values* tell the details of local folding states at residue resolution, thus being able to interpret the consequence of energetic frustrations introduced by nonnative hydrophobic inter-residue interactions.

A protein conformation can be interpreted with a two-dimensional $N \times N$ square distance matrix whose element $d_{ij}$ is the distance between residues *i* and *j*. This matrix is then transferred into the so-called contact-map when the element values replaced by 0 or 1, according to the following equation,

$$a_{ij} = \begin{cases} 1 & \text{if } d_{ij} < r_c \\ 0 & \text{else} \end{cases} \tag{4a}$$

where $r_c$ is the distance cutoff with a value between 7 to 15Å. Contact map is a simple two-dimensional representation of a protein's tertiary structure, thus it is particularly useful to monitor changes of secondary-structure elements through a large-scaled calculation. In the contact-map, a band along the main diagonal usually stands for an α-helix fragment whose contacts are formed among sequentially neighboring residues. On the other hand, β-sheet structures may be recognized as bands either parallel to (for parallel β-sheet) or perpendicular to the diagonal (for anti-parallel β-sheet), and the band thickness tells the number of strands within a β-sheet. Contact-map had been broadly used in analyzing secondary-structure features of TSEs in protein folding [69], structure modeling[70,71] and other molecular dynamics simulations[72]. Here in order to simplify the analysis, we modify Equation 4a as follows,

$$a_{ij} = \begin{cases} 1 & \text{if residue } i,j \text{ form a native contact} \\ 0 & \text{else} \end{cases},$$

and define contact-map of TSEs through ensemble average,

$$c_{ij} = \langle a_{ij} \rangle^{\text{TSE}}. \tag{4b}$$

Contact maps measure the probability that two residues tend to stay together as in their native states, which can be interpreted as a 2D projection of the tertiary structures.

**Results and Discussion**

We examined energetic frustrations in the all-*β* proteins fold by comparison study of the minimum model and frustrated model for 9 Ig-like *β*-sandwich homologous domains (Figure 1). These structures are mutually superposed with a small $C_\alpha$ root-mean-square-RMSD between 0.4 and 0.8Å (see Table 1), thus minimizing topological frustration differences and highlighting the effect of energetic frustrations caused by amino-acid replacements. Ig-like *β*-sandwich domains comprise of eight *β*-stands (B1-B8), of which the latter 6 ones formed 3 *β*-hairpins (BH1-BH3) (see Figure 2). The second *β*-stand B2 and the second *β*-hairpins BH2 comprise the first *β*-pleated sheet, and B1, BH1 and BH3 form the second *β*-sheet. These two *β*-sheets are tightly packed through linkers L2 at the bottom and L4 on the top. For clarity, we renamed each domain after the dominant hydrophilic-hydrophobic mutation (see the caption in Figure 2). Homotypic mutations exist in three structures: domain **R23A** has additional mutation of T21K, domain **N36I** has T47S, A57V, I77V and another hydrophilic-hydrophobic mutation Q1F, and domain **Y94D** has N55K, W98L. Mutations are divided into two groups: 1) the hydrophilic-to-hydrophobic mutation type: **R23A**, **N36I**, **D43A**, **T53A**, **D71A**, and 2) the hydrophobic-to-hydrophilic mutation type: **V60T**, **A80G**, **Y94D**. These systems are simulated using both the conventional coarse-grained Gō-like model and the frustrated model. To examine the perturbation of these mutations on protein folding, we calculated residual $\phi$-values and inter-residue contact maps, and thus determined the effect of energetic frustration as a change in residual $\phi$-value and inter-residue contact[73,74].

**Transition State Ensembles of Ig-like β-sandwich folds are sensitive to the energetic frustration mutation centers** The conformations in TSEs of Ig-like $\beta$-sandwich folds are sorted with respect to Q value — the total number of native contacts preserved in given conformation normalized by the total number of native contacts found in native states. Q value varies between 0 and 1, corresponding to the fully unfolded states and the fully folded native states, respectively. According to Boltzmann distribution theory, an apparent free energy of the system in terms of Q-value can be derived as the negative log of the conformation distribution probability $(-\ln P(Q))$. In VTF simulations, an averaged transition temperature $\bar{T}$ is used in determining the free energy temperature factor $k_\mathrm{B}\bar{T}$. Figure 3 compares the apparent "free energy" landscapes for the two models. The changes in free energy landscape are recognized as the shift of the location of central energy barrier, which corresponds to the TSEs. Essentially, two groups exist: one group

takes right-shift with the free energy profile shifting to the high Q-value end, indicating more native contacts preserved in TSEs. The second one takes a left-shit with less native contacts in TSEs. Five domains **HC19**, **R23A**, **D43A**, **D71A**, **Y94D** have a right-shifted free-energy change, a similar change had also been observed in the studies of Im9 all-$\alpha$ folds. However, unlike that in Im9 all-$\alpha$ folds, here in $\beta$-sandwich folds energetic frustrations seem not to increase the folding barriers. Instead, significant lower barriers are observed in domains **R23A**, **D43A**, indicating that in these structures TSEs are more stable with more native contacts upon energetic frustrations. Left-shift perturbations are found in domains **N36I**, **T53A**, **V60T** and **A80G**, indicating that in these domains energetic frustrations dissolve certain native contacts in TSEs. Slight decrease in energy barrier is observed in domain **N36I**, while **A80G** has a slight barrier increase. In the left-shift cases, the right end of the free energy profiles shift slightly to the left, indicating a few native contacts are destroyed in folded states. Taken together, energetic frustrations caused by remote nonnative-hydrophobic-contacts may either increase or decrease native contacts in TSEs of Ig-like β-sandwich folds, and the overall effect of energetic frustration tends to enhance the stability of TSEs for most of examined structures (except **A80G**), which is different from that of Im9 all-$\alpha$ domains where energetic frustration usually tends to destabilize TSEs.

**Energetic frustrations have highly heterogeneous effects on secondary folding of Ig-like β-sandwich folds**   The complexity of energetic frustration lies in the details of secondary-structures folding-status change, which can be detected by examining TSEs conformation distribution perturbation upon energetic frustration. Specifically, we calculated residual $\phi$-value changes upon the frustrations, and, considering that all examined structures share the same tertiary structure(see Table 1), these changes can be ascribed to the pure effects of energetic frustrations caused by the nonnative hydrophobic interactions. Particularly, since only a single hydrophilic-hydrophobic mutation exists between the examined structures (except domain **N36I**), and we can further associate the overall effects of energetic frustrations to the local environment where mutations happen.

Figure 4 shows residual $\phi$-value-changes derived from the two models. Due to the simplicity of the model and the vacuum environment of simulation, the absolute phi-values (see supplementary Figure S1) fall short of experimental measurements[51], and a comparison with experimental $\phi$-values is given in supplementary Figure S2. Compared with all-$\alpha$ Im9 domains, the calculations show that all-$\beta$ domains experience more complicated energetic frustrations, as revealed by the highly-irregular perturbations in their residual $\phi$-value distributions, No significant residual $\phi$-value change is found in both domains **N36I** and **T53A**, indicating the effects of energetic frustrations in these proteins are ignorable. For **V60T**, the residual $\phi$-values decrease in BH2 is balanced by increase in B1, leaving an averaged $\phi$-value almost unchanged. In **HC19,** significant $\phi$-value increases are found in linker L2, β-hairpins BH2 and BH3, while a small decrease happens in the first β-strand B1. **R23A** has detectable $\phi$-value increase in B1, BH3 and C-strand of L3. **D43A** has small increases in B1, L2, N-strand of BH1 and N-strand of BH2. **D71A** has larger $\phi$-value increases in B1 and detectable decreases in BH2. In **A80G**, residual $\phi$-values become smaller in B1 and BH3. **Y94D** obtains significant $\phi$-value increases in B1 and BH3 and small decrease in BH2. Taken together, both introduce (in hydrophilic-to-hydrophobic mutant) and elimination (in hydrophobic-to-hydrophilic mutant) of energetic frustrations result in complicated fluctuations to the folding of most secondary structures in *β*-sandwich structures.

It is interesting to compare energetic frustration effects in all *β*-sandwich domains with those in all-$\alpha$ Im9 proteins. Energetic frustration increase residual $\phi$-values for all examined all-*α* domains (see Table 2 in reference [54]), of which the largest increment happens in D51A whose mutation occurs in the center of the native-contact network. In this center area, a hydrophilic-to-hydrophobic mutation facilitates the formation of native-contacts among residues from the 3 surrounding long helices. However, in all *β*-sandwich structures, native contacts are much evenly distributed within the structures, and lack a center area where a relatively large energetic frustration may happen just like that in the all-*α* **D51A** domain[54]. Thus, energetic frustrations of *β*-sandwich structures show little dependence on the mutation locations.

The heterogeneity of energetic frustrations can be appreciated more clearly with changes of averaged $\phi$-values of secondary structures (Figure 5). The largest changes happen in B1 and BH3, the two elements occupy the N- and C-termini of the structure, respectively. These two secondary structures have a significant $\phi$-value increase for 5 of the 9 studied proteins and a detectable $\phi$-value decrease in **A80G** and **T53A**. The only inconsistence occurs in HC19 where B1 has lower $\phi$-values while BH3 has larger ones. Topologically, a good packing between B1 and BH3 should be critically important for *β*-sandwich structure to reach its native state. The second *β*-hairpin BH2 also has significant change in residual φ-value distribution: significant increase in **HC19**, **D43A**, **T53A** and **A80G** and moderate-to-weak decrease in other cases. More interestingly, BH2 seems to do the reverse as B1 and BH3 in altering their φ-values. For example, in **R23A**, **N36I**, **D71A**, **Y94D**, B1 and BH3 gain larger φ-values while BH2 has smaller ones. In **T53A**, **A80G**, smaller φ-values are found in B1 and BH3 while larger ones found in BH2. The second β-strand and the first β-hairpin show less φ-value perturbations as compared with **B1** and **BH2/3**. All linkers except **L2** have ignorable φ-value changes upon energetic frustration. Even for **L2** – the linker connects the two β-sheets at the bottom – only a moderate-to-weak change is observed. To summarize, the heterogeneity of energetic frustration in the folding of β-sandwich proteins lies in two folds: a single site-mutation might cause fundamentally different perturbation to the overall folding status (see the difference between **HC19** and **A80G**), indicating that energetic frustration is highly site-sensitive to the local environment; secondary structure elements may have different responses to the same energetic frustration as revealed by φ-value changes in BH2 and those in BH3 and B1.

**Energetic frustration alter the folding consistency between the folding patches in β-sandwich structures**    The enclosed topology structure of β-sandwich domains requires a concomitant formation of the secondary structure and then the tertiary structure in folding, resulted in a typical nucleation-condensation folding mechanism[47,75]. The folding contact-map of β-sandwich structures calculated from TSEs shows similar secondary-structure interactions during protein folding (see detailed comparison in supplementary Figure S3). The contact map in the upper left triangle represents energetic frustrated simulation, which is compared with that in the bottom right triangle derived from the conventional minimal model. The matrix diagonal records intensive interactions between intermediately neighboring anti-parallel *β*-strands as following: (B1, B2), (BH1, BH1), (BH2, BH2), (BH3, BH3). Moderate secondary-structure-contacts appear in the bottom right and upper left area, including anti-parallel *β*-strands (N-strand of BH1, N-strand of BH3), (B2, N-strand of BH2) and parallel *β*-strands (B1, C-strand of BH3). The strongest contacts happen in the N- and C-strands of BH3, especially at its central-turn. The second strongest interactions occur between BH1 and BH3 in the middle of the second *β*-sheet. The 9 proteins share similar contact maps highlighted by above-mentioned *β*-strands contact-patches. Thus they are likely to have similar overall TSEs during protein folding and share the same folding mechanism, which is consistent with previous studies in that immunoglobulin-like *β*-sandwich proteins tend to share a common nucleation-condensation folding pathway.

To examine the effects of energetic frustrations on the folding pathway of *β*-sandwich proteins, we calculated the contact-map differences by subtracting the upper triangle element from its lower triangle symmetric counterpart (see Figure 6). The color of matrix elements represents the contact-tendency between two residues: red for enhanced contact and blue for weakened contact. The heterogeneous effects of energetic frustrations are read from the inconsistent distributions of red and blue patches in the folding of 9 domains. The most prominent changes occur in *β*-strand pairs along the diagonal, including the three *β*-hairpins and the initial pair of (B1, B2). We noticed that in a single patch the color distribution is pretty even: they are either red or blue, indicating that collective folding/unfolding exists in the patches formed by neighboring *β*-strand pairs. In this sense, we call these *β*-strand pairs the folding patches.

Our calculations show that folding patches have different correlation with one another in response to energetic frustrations, depending on patch elements and the locations where mutations happen. For example, BH1 and BH2 (the two patches in the center of the diagonal) have the same folding propensity in **HC19** and **V60T**, but take the reverse tendencies in **N36I** and **Y94D**. Another two patches, B1-B2 and BH3, at the two ends of the diagonal, show a different response: both increase folding tendency in mutations **R23A**, **T53A**, **A80G** and **Y94D** but decrease folding tendency in **V60T**. The two elements in the patch pair, B1-B2 and B1-C-strand of BH3, show similar folding tendency correlation as does the patch pair B1-B2 and BH3. It is interesting to examine the site-mutation between **HC19** and **V60T**: energetic frustrations enhance the folding in all of the three *β*-hairpins in **HC19** but weaken the folding in **V60T**. These results suggest that energetic frustrations bring heterogeneous effects on the local folding status in *β*-sandwich structures. And at the same time, energetic frustrations generate alternative correlations between folding-patches, which are sensitive on local environment of mutations.

**Concluding remarks**

In this paper, we examined energetic frustrations in all-β structures by comparing the folding mechanisms of 9 homologous Ig-like β-sandwich domains. The studied domains share the same complex Greek-key topology and have similar tertiary structures, thus topological frustrations in their folding dynamics are minimized. Two sets of transition state ensembles were obtained through variable temperature simulations using the minimal and the frustrated Gō-like models, respectively. Energetic frustrations were modeled with non-native hydrophobic interactions using the standard Lennard-Jones potential function. The selected domains share the same sequence except a dominant single hydrophilic-hydrophobic mutation among them, thus a sequence-alignment arrangement might locate energetic frustrations in the 3D structures. Our calculations suggest that energetic frustrations cause highly heterogeneous effects on the folding of β-sandwich structures, as illustrated by a diverse change distribution in the folding of secondary structure elements and in correlations between folding-patches in neighboring patch-pairs. Our results also show that β-sandwich structure exhibits a high resistance to a variety of energetic frustrations in keeping its hydrophobic core in the nucleation-condensation folding.

Contrast to destabilizing the transition states in folding of compact all-$\alpha$ Im9 proteins[54], energetic frustrations tend to stabilize the transition states; this might be ascribed to the competition between the native hydrophobic contacts and nonnative hydrophobic contacts introduced by energetic frustrations. In this sense, energetic frustrations in β-sandwich structures also have a close interaction with the native-contact networks in reshaping protein-folding dynamics. Considering that both β-sandwich domains and all-$\alpha$ domains are all small and tightly-compact single-domain structures, their consistency in the interplay between energetic frustrations and native contact networks seems not difficult to appreciate[36]. However, the difference in secondary-structure components between these two types of structures makes energetic frustrations in β-sandwich proteins more heterogeneous, thus the formation of β-sheet in β-sandwich might feel more frustrated than does the $\alpha$-helix in all-$\alpha$ structures. Particularly, energetic frustrations due to single site mutation in linkers L3 and L2 cause large folding fluctuations in most secondary structures, and interestingly alter the folding correlations between different folding patches. These results suggest that protein folding in β-sandwich domains might be tuned by carefully manipulating energetic frustrations at residue level.

```
       ******************.* ************ **********:*******:*.*******************:**.************* *** ************
HC19   QAVVTQESALTTSPGETVTLTCRSSTGAVTTSNYANWVQEKPDHLFTGLIGGTNNRAPGVPARFSGSLIGDKAALTITGAQTEDEAIYFCALWYSNHWVFGGGTKLTVLG
R23A   QAVVTQESALTTSPGETVTLKCASSTGAVTTSNYANWVQEKPDHLFTGLIGGTNNRAPGVPARFSGSLIGDKAALTITGAQTEDEAIYFCALWYSNHWVFGGGTKLTVLG
N36I   FAVVTQESALTTSPGETVTLTCRSSTGAVTTSNYAIWVQEKPDHLFSGLIGGTNNRVPGVPARFSGSLIGDKAALTVTGAQTEDEAIYFCALWYSNHWVFGGGTKLTVLG
A80G   QAVVTQESALTTSPGETVTLTCRSSTGAVTTSNYANWVQEKPDHLFTGLIGGTNNRAPGVPARFSGSLIGDKAALTITGGQTEDEAIYFCALWYSNHWVFGGGTKLTVLG
Y94D   QAVVTQESALTTSPGETVTLTCRSSTGAVTTSNYANWVQEKPDHLFTGLIGGTNKRAPGVPARFSGSLIGDKAALTITGAQTEDEAIYFCALWDSNHLVFGGGTKLTVLG
```

Figure 1. Sequence alignment of the five selected Ig-like β-sandwich domains selected from SCOP. The abbreviations reads HC19: d1gigl1, R23A: d4a6yl1, N36I: d1etzl1, A80G: d2y06l1, Y94D: d3ks0l1. Additional four structures are derived from HC19 by artificial single-site mutation at site D43, T53, V60 and D71, and named as D43A, T53A, V60T, D71A, respectively.

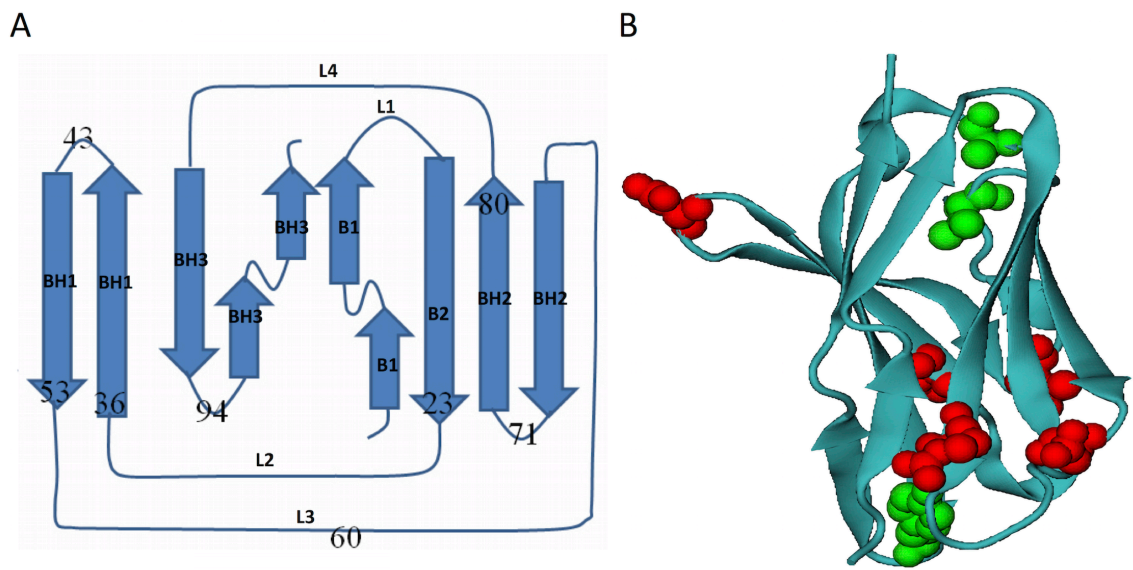

**Figure 2.** The topology of selected Ig-like β-sandwich domains and the location of mutations. (A) Topology of Ig-like β-sandwich domains. The secondary structures are β-strand **B1**: residue 1-12, linker **L1**: residue 13-15, **B2**: 16-24, **L2**: 25-35, β-hairpin **BH1**: 36-43 & 44-51, **L3**: 52-63, **BH2**: 64-70 & 71-80, **L4**: 81-85, and β-hairpin **BH3**: 86-96 & 97-110. (B) The tertiary structure of Ig-like β-sandwich domains with the mutations highlighted with solid spheres: read ones are hydrophilic-to-hydrophobic mutations while green ones are the reverse mutations. The is figure was prepared using VMD software[76].

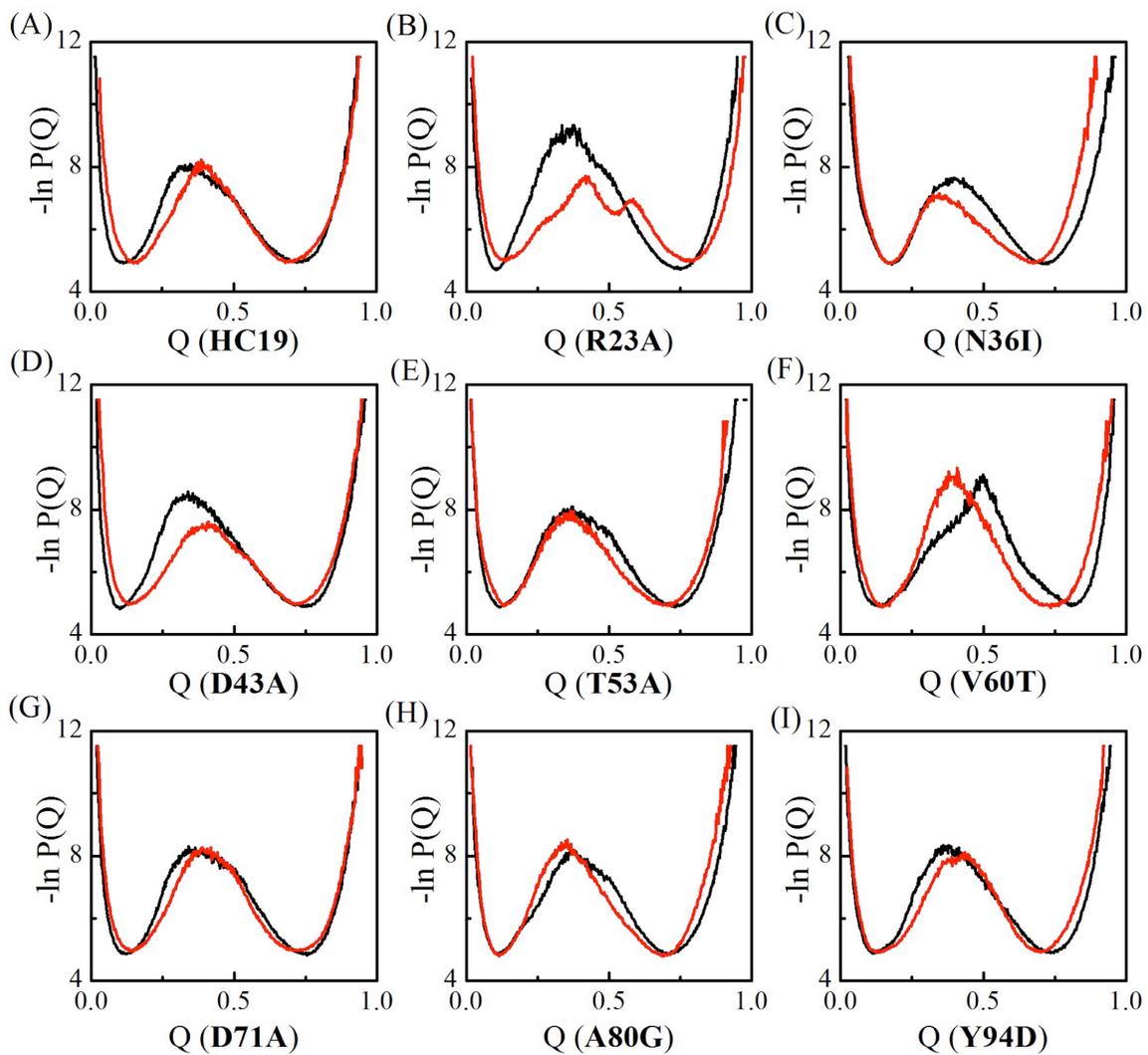

Figure 3. Apparent folding free energy changes for the 9 Ig-like β-sandwich domains. Black stands for the conventional Gō-like minimal model and red for frustration model.

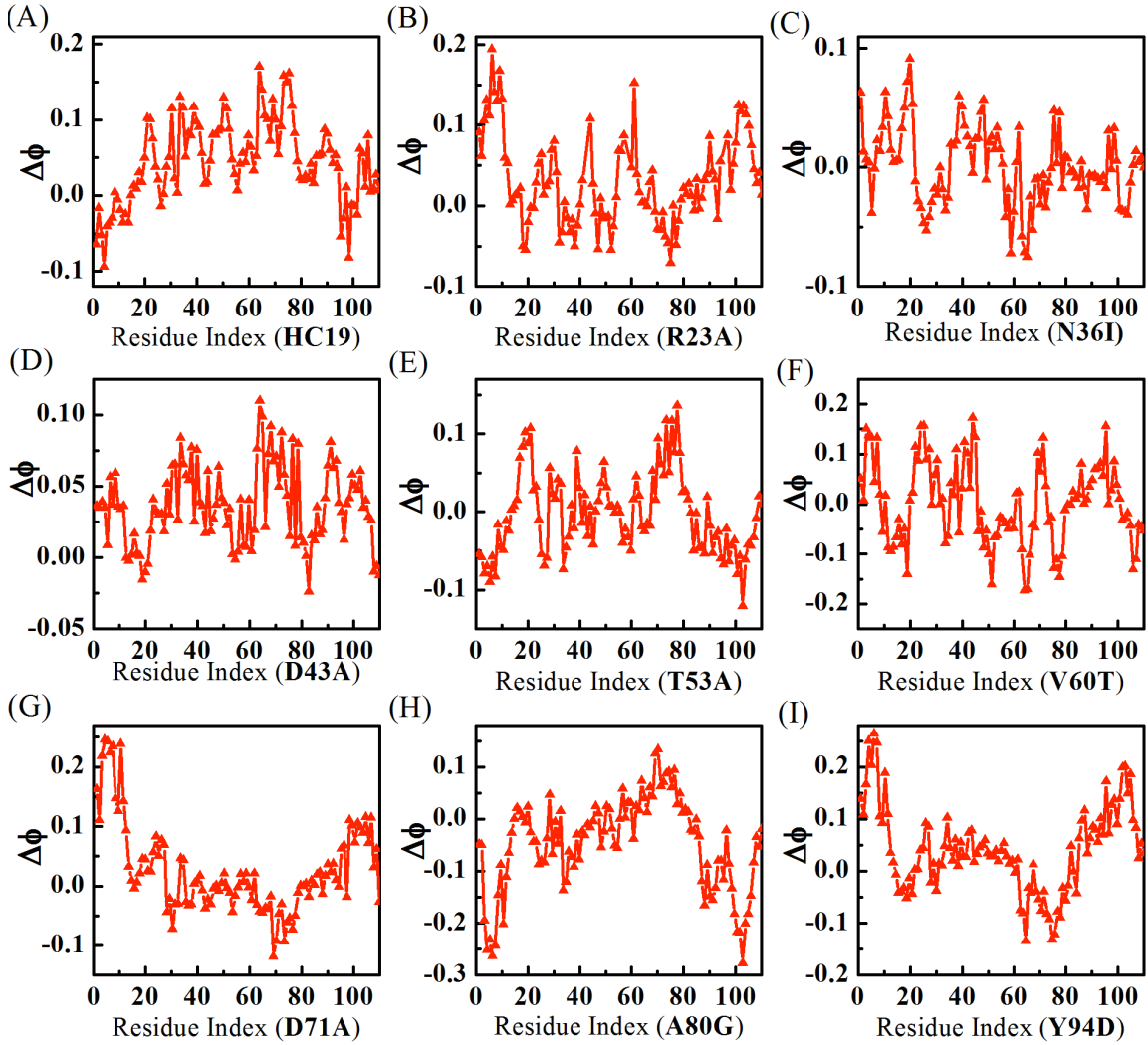

Figure 4. Energetic frustrations are heavily dependent on mutation locations in the β-sandwich structures.

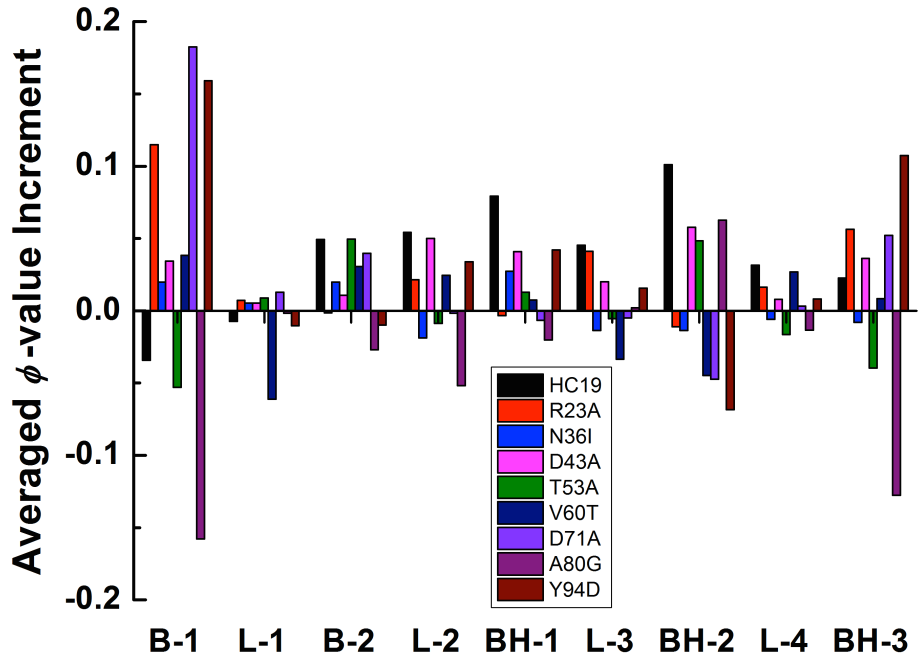

Figure 5. Averaged φ-value changes as a function of secondary structures for the nine β-sandwich domains.

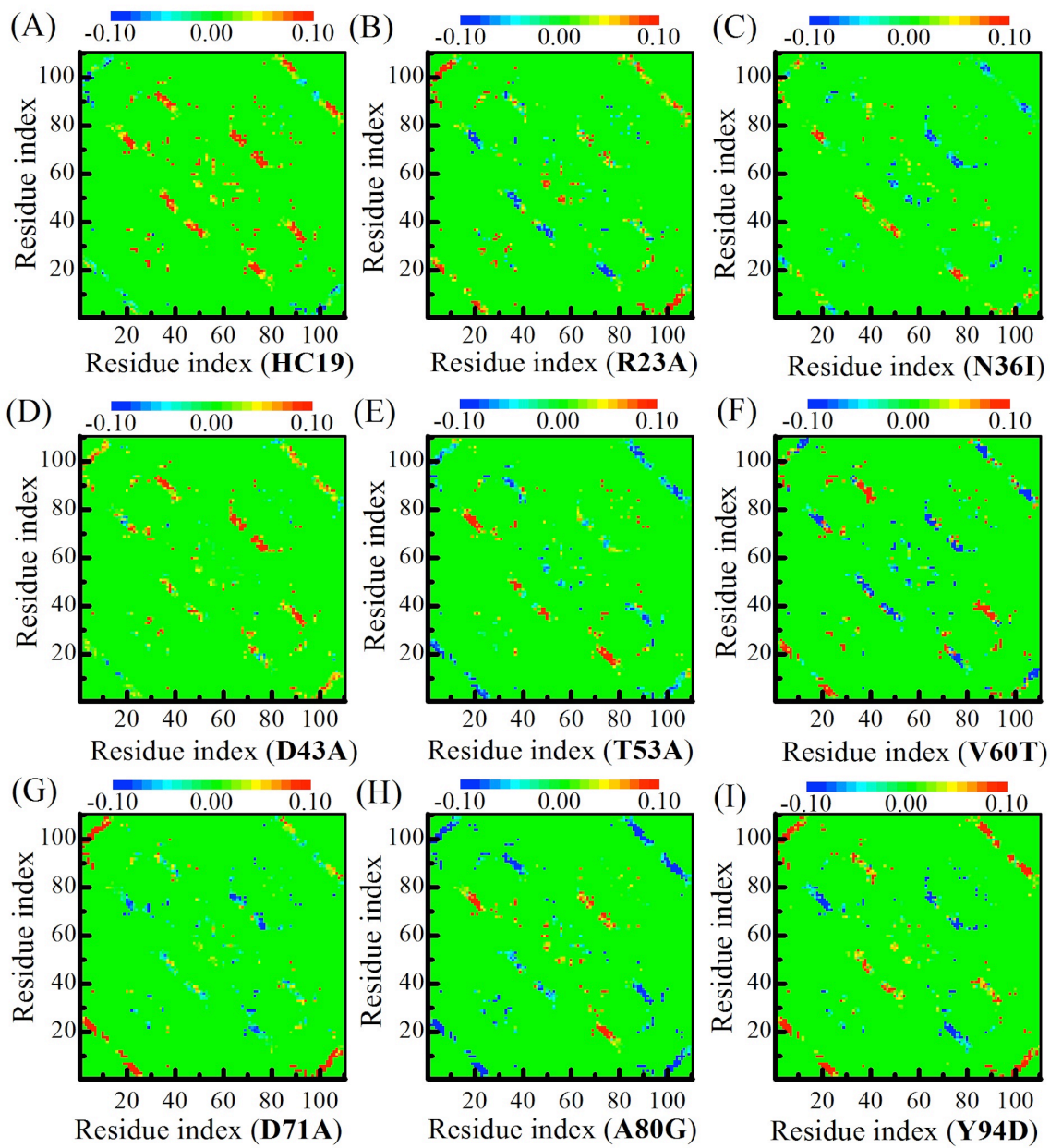

Figure 6. The contact difference maps for the nine β-sandwich domains. The difference of contact map is associated with the local environments of site-mutations.

**Table 1.** Backbone/$C_\alpha$ root mean square distance (in Å) between selected Ig-like $\beta$-sandwich domains.

|      | HC19      | R23A      | N36I      | A80G      | Y94D |
|------|-----------|-----------|-----------|-----------|------|
| **HC19** | 0         |           |           |           |      |
| **R23A** | 0.62/0.51 | 0         |           |           |      |
| **N36I** | 0.53/0.47 | 0.58/0.48 | 0         |           |      |
| **A80G** | 0.57/0.52 | 0.61/0.54 | 0.53/0.49 | 0         |      |
| **Y94D** | 0.76/0.79 | 0.77/0.73 | 0.65/0.62 | 0.69/0.64 | 0    |